\documentclass[twocolumn,english,showpacs,preprintnumbers,prb]{revtex4-1}
\usepackage[latin9]{inputenc}
\usepackage{amsmath}
\usepackage{amssymb}
\usepackage{graphicx}
\usepackage{esint}

\makeatletter

\usepackage{color}

\@ifundefined{textcolor}{}{%
 \definecolor{BLACK}{gray}{0}
 \definecolor{WHITE}{gray}{1}
 \definecolor{RED}{rgb}{1,0,0}
 \definecolor{GREEN}{rgb}{0,1,0}
 \definecolor{BLUE}{rgb}{0,0,1}
 \definecolor{CYAN}{cmyk}{1,0,0,0}
 \definecolor{MAGENTA}{cmyk}{0,1,0,0}
 \definecolor{YELLOW}{cmyk}{0,0,1,0}
}

\usepackage{color}

\usepackage{ulem}

\usepackage{aecompl}

\usepackage{epsfig}\usepackage{dcolumn}\usepackage{bm}

\usepackage{babel}

\makeatother

\usepackage{babel}
\begin{document}

\title{Unveiling Majorana Quasiparticles by a Quantum Phase Transition:
Proposal of a Current Switch}

\author{F. A. Dessotti$^{1}$, L. S. Ricco$^{1}$, Y. Marques$^{1}$, L. H. Guessi$^{2,3}$,\\
 M. Yoshida$^{2}$, M. S. Figueira$^{4}$, M. de Souza$^{2}$, Pasquale
Sodano$^{5,6}$, and A. C. Seridonio$^{1,2}$}
\affiliation{$^{1}$Departamento de F\'{i}sica e Qu\'{i}mica, Unesp - Univ Estadual
Paulista, 15385-000, Ilha Solteira, SP, Brazil\\
 $^{2}$IGCE, Unesp - Univ Estadual Paulista, Departamento de F\'{i}sica,
13506-900, Rio Claro, SP, Brazil\\
$^{3}$Instituto de F\'{i}sica de S\~{a}o Carlos, Universidade de S\~{a}o Paulo, C.P. 369, S\~{a}o Carlos, SP, 13560-970, Brazil\\
$^{4}$Instituto de F\'{i}sica, Universidade Federal Fluminense,
24210-340, Niterói, RJ, Brazil\\
 $^{5}$International Institute of Physics, Universidade Federal do
Rio Grande do Norte, 59078-400, Natal, RN, Brazil\\
 $^{6}$Departamento de F\'{i}sica Te\'{o}rica e Experimental, Universidade
Federal do Rio Grande do Norte, 59072-970 Natal, RN, Brazil}

\begin{abstract}
We propose a theoretical approach based on an interferometer composed
by two quantum dots asymmetrically coupled to isolated Majorana quasiparticles
(MQPs), lying on the edges of two topological Kitaev chains, respectively
via couplings $(t+\Delta)$ and $(\Delta-t)$. This setup enables
us to probe MQPs in a quite distinct way from the zero-bias peak feature.
Most importantly, the system behaves as a current switch made by two
distinct paths: (i) for the upper dot connected to both chains, the
device perceives both MQPs as an ordinary fermion and the current
crosses solely the lower dot, since current in the upper dot is prevented
due to the presence of the superconducting gap; and (ii) by suppressing
slightly the hybridization of the upper dot with one chain, the current
is abruptly switched to flow through this dot, once a trapped electron
as a bound state in the continuum (BIC) {[}Phys. Rev. B \textbf{93},
165116 (2016){]} appears in the lower dot. Such a current switch between
upper and lower dots characterizes the Quantum Phase Transition (QPT)
proposed here, being the ratio $t/\Delta$ the control parameter of
the transition. This QPT is associated with a change from an ordinary
fermionic excitation regime to a MQP in the interferometer, which
enables not only the fundamental revealing of MQPs, but also yields
a current switch assisted by them.
\end{abstract}


\maketitle

\section{Introduction}

A scenario of misinterpretations concerning the real existence of
a Majorana quasiparticle (MQP) in condensed matter systems\cite{Review1,Review2,Review3}
is due to the demand of the zero-bias peak (ZBP)\cite{wire1,wire2}
as the clear evidence for its proof, once such a characteristic may
stem from or masked by other phenomena\cite{PLee1,DSarma,PLee2} as
we will discuss below. Thereby in this work, we propose an alternative strategy to avoid this demand by introducing a novel based
MQP current switch by means of an emerging quantum phase transition
(QPT). Particularly, a MQP attached to an edge of a topological Kitaev chain\cite{Kitaev,UFMG,Nayana,Jelena,wire3,wire4,PSodano},
as known theoretically, has as fingerprint the fractional ZBP $G=0.5{e^{2}}/{h}$
appearing in the conductance through a quantum dot (QD)\cite{Baranger,Vernek}.
Such a signature is elusive, once other physical phenomena and experimental
difficulties can mask this feature by leading to a ZBP disregarding
the MQP picture. Thus some criticisms have been reported in the literature
addressing the validity of the ZBPs found in the experiments of Refs.\,{[}\onlinecite{wire1}{]}
and {[}\onlinecite{wire2}{]}, respectively for semiconducting nanowire
and magnetic adatom systems in the presence of strong spin-orbit and
magnetic fields with an \textit{s-}wave superconducting surface. In the experiments above, it is not clear whether the ZBPs are due to a genuine
isolated MQP or associated with disorder introduced by ordinary fermionic
states lying within the superconducting gap, crossed Andreev effect,
among others\cite{PLee1,DSarma,PLee2}. Furthermore, thermal broadening
together with a coherence length much longer than the Kitaev chain
size can also lead to the overlap of the MQPs at the chain edges,
thus suppressing the ZBP signature\cite{PLee2}.

\begin{figure}[!]
\includegraphics[width=0.47\textwidth]{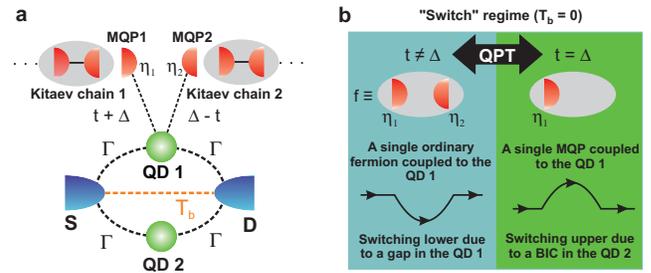} \protect\protect\protect\protect\protect\protect\protect\protect\protect\protect\protect\caption{\label{fig:PicZero} (Color online) (a) Double-QD interferometer connected
to two Kitaev chains with MQPs $\eta_{1}$ and $\eta_{2}$. The QDs
are hybridized via $\Gamma$ with the leads, while the upper QD with
the chains by means of $(t+\Delta)$ and $(\Delta-t)$, respectively.
$\mathcal{T}_{b}$ represents the background transmittance through
the source (S) and drain (D) leads. (b) QPT: the interferometer behaves
as an abrupt current switch for $\mathcal{T}_{b}=0$ when we tune
from the $t\protect\neq\Delta$ regime to $t=\Delta$.}
\end{figure}

Here our detection strategy adds a second QD to the
system developed in our previous work\cite{BICsM}, where a novel
technology for \textit{qubit} storage was proposed based on bound
states in the continuum (BICs)\cite{Neumann,SH,Capasso} formed by
MQPs. Thus we show that by using a double-QD interferometer connected
to a pair of topological Kitaev chains, as those sketched in Fig.\ref{fig:PicZero}(a),
that an emerging QPT then unveils a MQP when the setup operates under
the ``Switch\char`\"{} regime for the current, namely: if the upper
QD is coupled simultaneously to the Kitaev chains, the interferometer
``feels\char`\"{} the two MQPs at the edges of the chains just as
an ordinary fermion and the current travels exclusively through the
lower dot (blue panel of Fig.\ref{fig:PicZero}(b)) once the other
dot presents a superconducting gap that prevents the current. By reducing
slightly the coupling between one chain and the upper dot, then the
current suddenly switches, the QPT here addressed, to the path via
this dot (green panel of Fig.\ref{fig:PicZero}(b)) as aftermath of
the BIC rising in the lower dot that blocks the current through it,
which is a straight indication of the half-fermion nature imposed
by the unique MQP present in the system. Off the ``Switch'' regime
(see Fig.\ref{fig:Pic3}), the QPT persists still via the abrupt change
in the transmittance lineshape, but revealing novel fractional values
$G=0.25{e^{2}}/{h}$ and $G=0.75{e^{2}}/{h}$ for the MQPs.

Before starting the system analysis itself, we should call attention for the model validity within an experimental perspective, specially due to the role of disorder in realistic Kitaev chains. It is well known that \textit{p-}wave superconductors as aftermath of its spinless superconducting nature, which is accompanied by the topological phase, is then fragile against nonmagnetic elastic scattering. There are several reports in the literature covering the disorder issue in the case of Kitaev chains. As a matter of fact, as pointed out by Refs.\,{[}\onlinecite{PLee1},\onlinecite{DSarma}{]}, there is indeed the need of having high purity chains in order to observe topological superconductivity. In quantitative terms, one has Ref.\,{[}\onlinecite{DSarma}{]} pointing out that, if the ratio $\tau\gg\hbar J/(\Delta_{s}E_{SO})$ for the elastic scattering rate is fulfilled, the topological superconductivity should remain and as expected, the QPT we have found as well, wherein $J$ represents the exchange spin splitting, $\Delta_{s}$ the \textit{s-}wave pairing of the chain's host and $E_{SO}$ is the spin-orbit coupling energy. In general terms, the disorder is an important aspect, both on the theoretical framework\cite{Beenakker} as well as related to the fabrication of devices based on the Kitaev chains. Thus, in the present work we focus on the expected intrinsic effects of an ``ideal" Kitaev chain and for this reason, we do not carried out any analysis concerning the disorder issue. More specifically, as we are dealing with a phase transition at zero temperature, quasiparticle poisoning is thus ruled out\cite{Poisoning}.

\section{Theoretical System}

We employ an extension of the Hamiltonian inspired on the proposal
from Liu and Baranger, which is a spinless model to ensure topological
superconductivity\cite{Baranger}:
\begin{eqnarray}
\mathcal{H} & = & \underset{\alpha k}{\sum}\tilde{\varepsilon}_{\alpha k}c_{\alpha k}^{\dagger}c_{\alpha k}+\sum_{j=1}^{2}\varepsilon_{j}d_{j}^{\dagger}d_{j}+V\sum_{\alpha kj}(c_{\alpha k}^{\dagger}d_{j}+\text{{H.c.}})\nonumber \\
 & + & V_{SD}(\sum_{kp}c_{Sk}^{\dagger}c_{Dp}+\text{{H.c.}})+\frac{(t+\Delta)}{\sqrt{2}}(d_{1}-d_{1}^{\dagger})\eta_{1}\nonumber \\
 & + & i\frac{(\Delta-t)}{\sqrt{2}}(d_{1}+d_{1}^{\dagger})\eta_{2},\label{eq:TIAM}
\end{eqnarray}
where the electrons in the lead $\alpha=S,D$ are described by the
operator $c_{\alpha k}^{\dagger}$ ($c_{\alpha k}$) for the creation
(annihilation) of an electron in a quantum state $k$ with energy
$\tilde{\varepsilon}_{\alpha k}=\varepsilon_{k}-\mu_{\alpha}$, with
$\mu_{\alpha}$ as the chemical potential. For the QDs coupled to
leads, $d_{j}^{\dagger}$ ($d_{j}$) creates (annihilates) an electron
in the state $\varepsilon_{j}.$ $V$ (or $\Gamma=2\pi V^{2}\rho_{0},$
the Anderson parameter\cite{Anderson} with $\rho_{0}$ the metallic
leads density of states) stands for the hybridizations between the
QDs and the leads, which are considered equal to ensure the same conductance
through the source and drain leads, i.e., $G=G_{S}=G_{D}$\cite{Cao,Zheng}.
The QD 1 couples asymmetrically to the Kitaev chains with tunneling
amplitudes defined for convenience by $t_{L}\equiv(1/\sqrt{2})(t+\Delta)$ and $t_{R}\equiv(i/\sqrt{2})(\Delta-t),$ respectively
for the left and right MQPs $\eta_{1}=\eta_{1}^{\dagger}$ and $\eta_{2}=\eta_{2}^{\dagger}.$
We stress that such definitions constitute just a choice of gauge which allows us to catch the following phenomenology:
$t_{L}$ and $t_{R}$ change the last two terms of Eq.\,(\ref{eq:TIAM}) into $td_{1}f^{\dagger}+\Delta f^{\dagger}d_{1}^{\dagger}+\text{{H.c.}},$
when the ordinary fermion substitution $f=\frac{1}{\sqrt{2}}(\eta_{1}+i\eta_{2})$
and $f^{\dagger}=\frac{1}{\sqrt{2}}(\eta_{1}-i\eta_{2})$ into Eq.\,(\ref{eq:TIAM}) is adopted. As a result,
electrons within $f$ and $d_{1}$ beyond the normal tunneling $t$ between them, become bounded as a Cooper pair
with binding energy $\Delta.$  It is worth mentioning that $\Delta$ used here refers to the coupling term as indicated in Fig.\ref{fig:PicZero}(a) and it does not represent the superconducting gap of the \textit{p-}wave Kitaev chains. However, the emergence of such a parameter can be understood as consequence of the proximity effect arising from the \textit{s-}wave superconductors hosting the Kitaev chains. We would like to clarify that if the actual gauge imposed above were another, the QPT as well as the electric current switch feature would be triggered by fixing $t_{L}$ and increasing slightly, for instance, $t_{R}$ (or vice-versa). Such a tuning of the amplitudes $t_{L}$ and $t_{R}$ nowadays is completely possible, thus turning the feasibility of the proposal concrete experimentally.

\subsection{Green's functions }
To calculate the conductance $G$, which depends upon the retarded
Green's functions $\tilde{\mathcal{G}}_{\mathcal{AB}}$ in the energy
domain $\varepsilon,$ with $\mathcal{A}$ and $\mathcal{B}$ as fermionic
operators belonging to the Hamiltonian $\mathcal{\mathcal{H}},$ we
should employ the equation-of-motion (EOM) method \cite{book} summarized
as follows $\omega\tilde{\mathcal{G}}_{\mathcal{AB}}=(\varepsilon+i0^{+})\tilde{\mathcal{G}}_{\mathcal{AB}}=[\mathcal{A},\mathcal{B^{\dagger}}]_{+}+\tilde{\mathcal{G}}_{\left[\mathcal{A},\mathcal{\mathcal{H}}_{i}\right]\mathcal{B}}.$
By applying the EOM on $\mathcal{G}_{d_{j}d_{l}}=-\frac{i}{\hbar}\theta\left(\tau\right){\tt Tr}\{\varrho[d_{j}\left(\tau\right),d_{l}^{\dagger}\left(0\right)]_{+}\}$
here expressed in terms of the density matrix $\varrho$ for Eq.\,(\ref{eq:TIAM})
and the Heaviside function $\theta\left(\tau\right)$, we change to
the energy domain $\varepsilon$ and obtain the following relation:
\begin{equation}
(\varepsilon-\varepsilon_{j}-\Sigma-\delta_{j1}\Sigma_{\text{{MQPs}}})\tilde{\mathcal{G}}_{d_{j}d_{l}}=\delta_{jl}+\Sigma\sum_{\tilde{l}\neq j}\tilde{\mathcal{G}}_{d_{\tilde{l}}d_{l}},\label{eq:GFs}
\end{equation}
with $\Sigma=-(\sqrt{x}+i)(1+x)^{-1}\Gamma,$ $x=(\pi\rho_{0}V_{SD})^{2},$
the MQPs self-energy $\Sigma_{\text{{MQPs}}}=K(t,\Delta)+(2t\Delta)^{2}K\tilde{K},$
where $K(t,\Delta)=[\varepsilon^{2}+2i\varepsilon0^{+}-(0^{+})^{2}]^{-1}\omega(t^{2}+\Delta^{2}),$
$K=[\varepsilon^{2}+2i\varepsilon0^{+}-(0^{+})^{2}]^{-1}\omega,$
$\tilde{K}=[\varepsilon+\varepsilon_{1}+\bar{\Sigma}-K(t,\Delta)]^{-1}K$
and $\bar{\Sigma}$ as the complex conjugate of $\Sigma.$ Thus the
solution of $\tilde{\mathcal{G}}_{d_{j}d_{l}}$ provides

\begin{equation}
\tilde{\mathcal{G}}_{d_{1}d_{1}}=\frac{1}{\varepsilon-\varepsilon_{1}-\Sigma-\Sigma_{\text{{MQPs}}}-\mathcal{C}_{2}}\label{eq:GF1}
\end{equation}
as the Green's function of the QD 1, with $\mathcal{C}_{j}=\Sigma^{2}(\varepsilon-\varepsilon_{j}-\Sigma)^{-1}$
as the self-energy due to the presence of the \textit{$j^{th}$} QD.
In the case of the QD 2, we have
\begin{equation}
\tilde{\mathcal{G}}_{d_{2}d_{2}}=\frac{1-\tilde{\mathcal{G}}_{d_{1}d_{1}}^{0}\Sigma_{\text{{MQPs}}}}{\varepsilon-\varepsilon_{2}-\Sigma-\frac{\tilde{\mathcal{G}}_{d_{1}d_{1}}^{0}}{\tilde{\mathcal{G}}_{d_{2}d_{2}}^{0}}\Sigma_{\text{{MQPs}}}-\mathcal{C}_{1}},\label{eq:GF2}
\end{equation}
where $\tilde{\mathcal{G}}_{d_{1}d_{1}}^{0}=(\varepsilon-\varepsilon_{1}-\Sigma)^{-1}$
and $\tilde{\mathcal{G}}_{d_{2}d_{2}}^{0}=(\varepsilon-\varepsilon_{2}-\Sigma)^{-1}$
represent the corresponding Green's functions for the single QD system
without MQPs. The mixed Green's functions are $\tilde{\mathcal{G}}_{d_{2}d_{1}}=\Sigma(\varepsilon-\varepsilon_{2}-\Sigma)^{-1}\tilde{\mathcal{G}}_{d_{1}d_{1}}$
and $\tilde{\mathcal{G}}_{d_{1}d_{2}}=\Sigma(\varepsilon-\varepsilon_{1}-\Sigma-\Sigma_{\text{{MQPs}}})^{-1}\tilde{\mathcal{G}}_{d_{2}d_{2}}.$

\subsection{Conductance }
In what follows we derive the Landauer-Büttiker formula for the zero-bias
conductance $G=(e^{2}/h)\mathcal{T}\left(\varepsilon=0,t,\Delta\right)$
at temperature $T=0$\cite{book}. Such a quantity is a function of
the transmittance $\mathcal{T}\left(\varepsilon,t,\Delta\right)$
as follows:
\begin{align}
\mathcal{T}\left(\varepsilon,t,\Delta\right) & =\mathcal{T}_{b}+2\sqrt{\mathcal{T}_{b}\mathcal{R}_{b}}\tilde{\Gamma}\sum_{j\tilde{j}}{\tt Re}\{\tilde{\mathcal{G}}_{d_{j}d_{\tilde{j}}}\left(\varepsilon\right)\}\nonumber \\
 & -(1-2\mathcal{T}_{b})\tilde{\Gamma}\sum_{j\tilde{j}}{\tt Im}\{\tilde{\mathcal{G}}_{d_{j}d_{\tilde{j}}}\left(\varepsilon\right)\}\nonumber \\
 & =\mathcal{T}_{b}+\sum_{j}\mathcal{T}_{jj}\left(\varepsilon,t,\Delta\right)+\sum_{j}\mathcal{T}_{j\bar{j}}\left(\varepsilon,t,\Delta\right),\label{eq:trans2}
\end{align}
where $j,\tilde{j}=1,2$ and $\bar{j}=1,2$ respectively for $j=2,1$
as labels to correlate distinct QDs, $\tilde{\Gamma}=\frac{\Gamma}{1+x}$
is an effective dot-lead coupling, $\mathcal{T}_{b}=\frac{4x}{\left(1+x\right)^{2}}$
represents the background transmittance and $\mathcal{R}_{b}=1-\mathcal{T}_{b}=\frac{\left(1-x\right)^{2}}{\left(1+x\right)^{2}}$
is the corresponding reflectance, both in the absence of the QDs and
MQPs, $\mathcal{T}_{jj}\left(\varepsilon,t,\Delta\right)$ gives the
transmittance through the $j^{th}$ QD, while the crossed term $\mathcal{T}_{j\bar{j}}\left(\varepsilon,t,\Delta\right)$
accounts for interference processes between these dots. For $t=\Delta$, we recover the Green's functions derived in Refs.\,{[}\onlinecite{JAP1}{]} and {[}\onlinecite{JAP2}{]}. In such works, we point out that the profiles of $\mathcal{T}\left(\varepsilon,t=\Delta\right)$ as a function of the single particle energy $\varepsilon$ for a given symmetric detuning $\Delta\varepsilon$ of the QDs appear analyzed in great detail, since this case corresponds to the unique presence of a Kitaev chain. Thus, a robust Majorana ZBP is found when $\mathcal{T}_{b}=0$ and the corresponding dip for $\mathcal{T}_{b}=1$ is observed. We then suggest the reader to see Refs.\,{[}\onlinecite{JAP1}{]} and {[}\onlinecite{JAP2}{]} where $\mathcal{T}\left(\varepsilon,t=\Delta\right)$ versus $\varepsilon$ for finite values of $\Delta\varepsilon$ can be found. In the limit of $t\neq\Delta,$ we have checked that the Majorana ZB anomaly (peak or dip) is not verified. For this latter case, see for instance Fig.\ref{fig:Pic1}(c), wherein $\mathcal{T}_{11}\left(\varepsilon,t\neq\Delta\right)$ and $\mathcal{T}_{21}\left(\varepsilon,t\neq\Delta\right)+\mathcal{T}_{12}\left(\varepsilon,t\neq\Delta\right)$ as functions of $\varepsilon$ appear explicitly for several $\Delta\varepsilon.$ Additionally, it is worth noticing that the QPT as well as the electric current switch feature of the transmittance, just emerge when $\varepsilon=0$ and as a function of $\Delta\varepsilon.$

\section{Results and Discussion }
Based on the Green's functions derived previously together with Eq.(\ref{eq:trans2})
for the total transmittance $\mathcal{T}\left(\varepsilon,t,\Delta\right)$,
we focus on two cases ruled by analytical expressions here determined
for $\mathcal{T}\left(\varepsilon=0,t\neq\Delta,\Delta\varepsilon\right)$
and $\mathcal{T}\left(\varepsilon=0,t=\Delta,\Delta\varepsilon\right)$,
as follows: the ``Switch\char`\"{} regime for the current, sketched
in details in Figs.\ref{fig:Pic1}(a) and \ref{fig:Pic2}(a), which
rises when we set the device to $\mathcal{T}_{b}=0,$ $\varepsilon=0$
(zero-bias) with $\varepsilon_{1}=\frac{\Delta\varepsilon}{2}$ and
$\varepsilon_{2}=-\frac{\Delta\varepsilon}{2}$ as the symmetric detuning
for the QDs; Off the ``Switch\char`\"{} regime, where just $\mathcal{T}_{b}=1$
is changed.

It is worth mentioning that the control parameter of the QPT is given
by the ratio $t/\Delta.$ Thus from now on, in Figs.\ref{fig:Pic1},
\ref{fig:Pic2} and \ref{fig:Pic3} we identify by the label ``numerical\char`\"{}
those curves determined by Eq.(\ref{eq:trans2}) by considering $t=6\Gamma$
with $\Delta=7\Gamma$ and $t=\Delta=6\Gamma$ respectively for the
limits $t\neq\Delta$ and $t=\Delta,$ otherwise the analytical expressions
for $\mathcal{T}\left(\varepsilon=0,t\neq\Delta,\Delta\varepsilon\right)$
and $\mathcal{T}\left(\varepsilon=0,t=\Delta,\Delta\varepsilon\right)$
here obtained appear with the label ``analytical\char`\"{}, which
are functions indeed explicitly independent on $t$ and $\Delta.$ Here the energy scale
adopted in the simulations is the Anderson parameter $\Gamma$\cite{Anderson}.

\subsection{Sharpness of the switching action}
We highlight that the analytical expressions that we have found, which
will appear later on, in the limits $t\neq\Delta$ and $t=\Delta,$
wondrously reveals a universal behavior by means of the independence
on the parameters $t$ and $\Delta.$ Consequently, it means that
numerically speaking, i.e., without performing the aforementioned
analytical simplifications in Eq.(\ref{eq:trans2}) for $\mathcal{T}\left(\varepsilon,t,\Delta\right)$,
the emulation of a weak suppression of the coupling between the QD
1 and Kitaev chain 2, can be realized by a slight change in $\Delta$
by fixing $t$ within such an equation, which then makes the system
to undergo a QPT, since the phases $t\neq\Delta$ and $t=\Delta$
are not smoothly connected being characterized by an absence of a
crossover.

As a result, there is a sudden change in the transmittance profile
as we will see, in which each phase is recognized by the derived analytical
expression describing $\mathcal{T}\left(\varepsilon=0,t\neq\Delta,\Delta\varepsilon\right)$
and $\mathcal{T}\left(\varepsilon=0,t=\Delta,\Delta\varepsilon\right),$ respectively. It means that, the sharpness of the switching action of  $\mathcal{T}\left(\varepsilon=0,t,\Delta,\Delta\varepsilon\right)$ is given by a Dirac delta behavior pinned at $t=\Delta$ when $t-\Delta$ is varied.

\subsection{The ``Switch\char`\"{} regime }
\begin{figure}[!]
\includegraphics[width=0.5\textwidth]{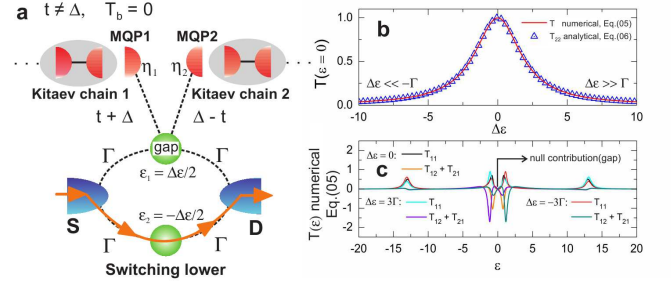} \protect\protect\protect\protect\protect\protect\protect\protect\protect\protect\protect\caption{\label{fig:Pic1} (Color online) The ``Switch\char`\"{} regime $\mathcal{T}_{b}=0$:
(a) The QD 1 perceives the two MQPs $\eta_{1}$ and $\eta_{2}$ (an
ordinary fermion), in particular with $t\protect\neq\Delta.$ (b)
As a result of (a), just the zero-bias transmittance ($\varepsilon=0$)
$\mathcal{T}_{22}$ contributes to the total system transmittance
$\mathcal{T}$ as a function of the symmetric detuning $\Delta\varepsilon$
for the QDs. (c) The transmittances $\mathcal{T}_{11}$ and $\mathcal{T}_{12}+\mathcal{T}_{21}$
exhibit a superconducting gap to prevent the current.}
\end{figure}

By considering $\mathcal{T}_{b}=0$ and $t\neq\Delta,$ we mimic both
the Kitaev chains of Fig.\ref{fig:Pic1} coupled to the QD 1, then
allowing the current to switch through the lower QD only, where we
see diagrammatically by the orange arrows, the current crossing solely
the QD 2 of panel (a) of the same figure. Within this regime, the
total transmittance reduces to

\begin{equation}
\mathcal{T}\left(\varepsilon=0,t\neq\Delta,\Delta\varepsilon\right)=\mathcal{T}_{22}=\frac{\Gamma^{2}}{\Gamma^{2}+\frac{1}{4}\Delta\varepsilon^{2}}\label{eq:T22}
\end{equation}
as outlined in Fig.\ref{fig:Pic1}(b), where we recognize $\mathcal{T}_{22}=1$
for the QDs on resonance ($\Delta\varepsilon=0)$ that connect the
metallic leads via the Fermi energy ($\varepsilon=0$) through the
QD 2 and $\mathcal{T}_{22}=0,$ the off-resonance case ($\Delta\varepsilon\gg\Gamma$
and $\Delta\varepsilon\ll-\Gamma$) wherein such a connection is found
truncated.

The unique contribution from $\mathcal{T}_{22}$ to the total transmittance
of the system, then lies on the features $\mathcal{T}_{11}\left(\varepsilon=0,t\neq\Delta,\Delta\varepsilon\right)=0$
and $\mathcal{T}_{12}\left(\varepsilon=0,t\neq\Delta,\Delta\varepsilon\right)+\mathcal{T}_{21}\left(\varepsilon=0,t\neq\Delta,\Delta\varepsilon\right)=0$
that do ensure a null contribution to $\mathcal{T}\left(\varepsilon=0,t\neq\Delta,\Delta\varepsilon\right)$
as can be verified in Fig.\ref{fig:Pic1}(c) as a function of $\varepsilon,$
for different $\Delta\varepsilon$ values.

The partial transmittance $\mathcal{T}_{11}\left(\varepsilon=0,t\neq\Delta,\Delta\varepsilon\right)=0$
reflects that the QD 1 perceives both the two MQPs at the edges of
the Kitaev chains as an ordinary fermionic zero mode, in such a way
that the splitting of this mode occurs in $\mathcal{T}_{11},$ which
opens a superconducting gap as pointed out in Fig.\ref{fig:Pic1}(c),
which prevents the current as a result. On the other hand, $\mathcal{T}_{12}\left(\varepsilon=0,t\neq\Delta,\Delta\varepsilon\right)+\mathcal{T}_{21}\left(\varepsilon=0,t\neq\Delta,\Delta\varepsilon\right)=0$
encode the scattering of electrons traveling forth and back between
the upper and lower QDs, which are phase shifted by $\pi,$ i.e.,
$\mathcal{T}_{12}\left(\varepsilon=0,t\neq\Delta,\Delta\varepsilon\right)=-\mathcal{T}_{21}\left(\varepsilon=0,t\neq\Delta,\Delta\varepsilon\right)$
canceling the net current through this path.

\begin{figure}[!]
\includegraphics[width=0.5\textwidth]{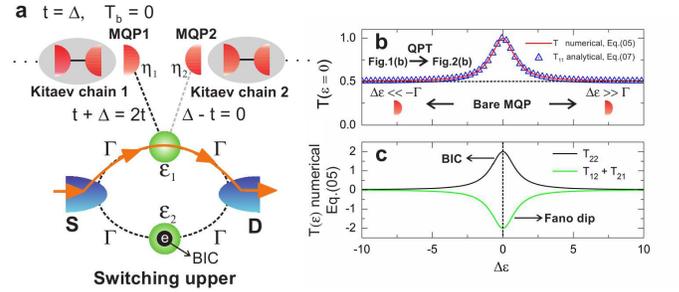} \protect\protect\protect\protect\protect\protect\protect\protect\protect\protect\protect\protect\caption{\label{fig:Pic2}(Color online) The ``Switch\char`\"{} regime $\mathcal{T}_{b}=0$:
(a) The current is switched upper (orange arrows) through the QD 1,
when such a dot perceives solely the MQP $\eta_{1}$ for $t=\Delta.$
(b) A QPT occurs due to the system abrupt change from $t\protect\neq\Delta$
to $t=\Delta,$ giving rise to a transmittance profile fully distinct
with respect to that found in Fig.\ref{fig:Pic1}(b). (c) In this
case, the transmittance $\mathcal{T}_{22}$ is canceled by a Fano
dip in $\mathcal{T}_{12}+\mathcal{T}_{21}.$ An electron $e$ is trapped
within the lower QD as a BIC.}
\end{figure}

The case $t=\Delta$ switches the upper QD suddenly, thus characterizing
the QPT found in panel (a) of Fig.\ref{fig:Pic2}, then yielding solely
\begin{equation}
\mathcal{T}\left(\varepsilon=0,t=\Delta,\Delta\varepsilon\right)=\mathcal{T}_{11}=\frac{1}{2}+\frac{1}{2}\frac{\Gamma^{2}}{\Gamma^{2}+\Delta\varepsilon^{2}}\label{eq:T11}
\end{equation}
as it appears in Fig.\ref{fig:Pic2}(b), which differently from Fig.\ref{fig:Pic1}(b),
gives $\mathcal{T}_{11}=0.5$ in the limits $\Delta\varepsilon\gg\Gamma$
and $\Delta\varepsilon\ll-\Gamma$: see the half-spheres outlined
in the panel (b) of Fig.\ref{fig:Pic2} for such a pictorial representation,
which points out the zero mode of the MQP at the edge in the Kitaev
chain 1 that leaked into the QD 1\cite{Vernek}. In such a case, the
path via $\mathcal{T}_{11}$ is chosen due to the emerging Fano dip\cite{Fano1,Fano2}
in $\mathcal{T}_{12}\left(\varepsilon=0,t=\Delta,\Delta\varepsilon\right)+\mathcal{T}_{21}\left(\varepsilon=0,t=\Delta,\Delta\varepsilon\right)$
that results in a perfect destructive interference with the resonance
lineshape of $\mathcal{T}_{22}\left(\varepsilon=0,t=\Delta,\Delta\varepsilon\right)$
as it appears in Fig.\ref{fig:Pic2}(c), which cancels the possibility
of other paths for the current through the system. This perfect cancelation
points out that an electron is trapped within the lower QD as a BIC,
which then blocks the current to cross from the source towards drain.
We highlight that the Fano interference mechanism of the BIC emergence
can be found discussed in great detail in Refs.\,{[}\onlinecite{BIC1}{]}
and {[}\onlinecite{BICsG}{]} for the graphene system. Particularly,
the BIC phenomenon here verified occurs when the nature of the device
is purely due to a MQP.

Thus, distinctly from the situation $t\neq\Delta$ for an ordinary
fermion present, where the absence of states in QD 1 due to a gap
that prevents the current flow, in the $t=\Delta$ regime a single
electron is bounded to the QD 2, which does not allow extra electrons
to pass as aftermath of the electronic state of this QD, which is
filled indefinitely due to the BIC phenomenon that puts into such
a dot an electron with an infinite lifetime.

\subsection{Off the ``Switch\char`\"{} regime }
\begin{figure}
\includegraphics[width=0.5\textwidth]{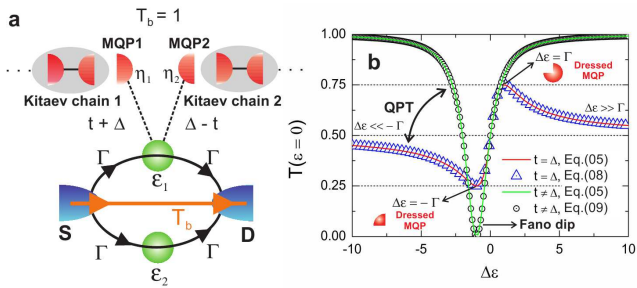} \protect\protect\protect\protect\protect\protect\protect\protect\protect\protect\protect\protect\caption{\label{fig:Pic3}(Color online) Off the ``Switch\char`\"{} regime
$\mathcal{T}_{b}=1$: (a) The lead-lead coupling $\mathcal{T}_{b}$
now exists and the current crosses the QDs as well as the middle region
between the leads. (b) Even without the current switch feature in
the device, the QPT remains made explicit by the abrupt change in
the transmittance profile, wherein novel MQP signatures rise given
by $\mathcal{T}=0.25$ and $\mathcal{T}=0.75,$ respectively for $\Delta\varepsilon=-\Gamma$
and $\Delta\varepsilon=\Gamma.$}
\end{figure}

For this regime we have $\mathcal{T}_{b}=1$ and the absence of the
current switch feature as depicted in Fig.\ref{fig:Pic3}(a), where
the current crosses simultaneously the QDs and the middle region between
the metallic leads, once we have checked that all terms of Eq.(\ref{eq:trans2})
are relevant to the total transmittance. Here Fig.\ref{fig:Pic3}(b)
exhibits the QPT wherein an abrupt change in the transmittance profile
still remains, i.e.,

\begin{equation}
\mathcal{T}\left(\varepsilon=0,t=\Delta,\Delta\varepsilon\right)=\frac{1}{2}+\frac{1}{2}\frac{\Delta\varepsilon\Gamma}{\Gamma^{2}+\Delta\varepsilon^{2}}\label{eq:T1}
\end{equation}
and
\begin{equation}
\mathcal{T}\left(\varepsilon=0,t\neq\Delta,\Delta\varepsilon\right)=1-\frac{\Gamma^{2}}{(\Gamma+\Delta\varepsilon)^{2}+\Delta\varepsilon^{2}},\label{eq:T2}
\end{equation}
due to the change from the case $t=\Delta$ to $t\neq\Delta.$ The
latter function describes the situation in which the QD 1 perceives
the Kitaev chains 1 and 2 as an ordinary fermion, being characterized
by $\mathcal{T}=1$ when the dots are found very far from the resonance
($\Delta\varepsilon\gg\Gamma$ and $\Delta\varepsilon\ll-\Gamma$)
recovering the background transmittance $\mathcal{T}_{b}=1$ of the
leads, in addition to the point $\Delta\varepsilon=-\Gamma$ where
$\mathcal{T}=0$ giving rise to a perfect Fano destructive interference.

By entering into $t=\Delta$ regime, the transmittance alters drastically
its profile thus making explicitly the QPT as a result of the isolated
MQP 1, as can be verified in the panel (b) of the same figure.

It is worth mentioning that, in particular, we predict well-established
fractional values for the transmittance. From the expression $\mathcal{T}\left(\varepsilon=0,t=\Delta,\Delta\varepsilon\right)$
above, one can show that especially for the detuning of the QDs $\Delta\varepsilon=\pm\Gamma,$
we find $\mathcal{T}\left(\varepsilon=0,t=\Delta,\Delta\varepsilon=-\Gamma\right)=0.25$
and $\mathcal{T}\left(\varepsilon=0,t=\Delta,\Delta\varepsilon=\Gamma\right)=0.75,$
which point out ``dressed\char`\"{} MQPs traveling through the interferometer,
appearing depicted by the incomplete spheres in Fig.\ref{fig:Pic3}(b)
in order to indicate pictorially the emergence of such quasiparticles.
These novel fractional values for the transmittance exactly placed
at $\Delta\varepsilon=-\Gamma$ and $\Delta\varepsilon=\Gamma$ are
then helpful to recognize that one isolated MQP lies on the Kitaev
chain 1.

\section{Conclusions }
In summary, by assuming $\mathcal{T}_{b}=0$ we have the phase $t\neq\Delta$
with $\mathcal{T}=\mathcal{T}_{22}$ (switching lower) when the QD
1 ``feels\char`\"{} both the MQPs at the edges of the Kitaev chains
as an ordinary fermion, followed by the phase $t=\Delta$ wherein
$\mathcal{T}=\mathcal{T}_{11}$ (switching upper) for this dot solely
connected to the Kitaev chain 1, when the system has a single MQP
isolated. Hence, the sudden change in the path for the current, which
does not have nothing to do with the elusive ZBP signature, is then
triggered by the QPT here addressed and serves not only to reveal
a MQP isolated in the system, but also to propose as an application
a current switch assisted by MQPs. Furthermore for $\mathcal{T}_{b}=1$,
we have found the novel fractional values $G=0.25{e^{2}}/{h}$ and
$G=0.75{e^{2}}/{h}$ in the conductance when the system is off the
``Switch\char`\"{} regime, which can help to recognize a MQP.

\section{Acknowledgments }
This work was supported by CNPq, CAPES, 2014/14143-0, 2015/23539-8, 2015/26655-9
S{ã}o Paulo Research Foundation (FAPESP).

\end{document}